\documentclass{article}
\usepackage{spconf,amsmath,graphicx}
\usepackage{todonotes}
\usepackage{caption}
\usepackage{subcaption}
\usepackage{multirow}
\usepackage{amsthm}
\usepackage{graphicx}
\newtheorem{definition}{Definition}
\usepackage[titletoc, title, page]{appendix}
\usepackage{algorithm}
\usepackage[noend]{algpseudocode}
\usepackage{booktabs}
\usepackage{times}
\usepackage{textcomp}
\usepackage{mathtools}
\usepackage{amsfonts}

\newcommand{\sci}[2]{#1 \mathrm{e}\protect\scalebox{0.9}{#2}} 
\newcommand{\Input}[1]{\Statex \textit{Input}: \textit{\footnotesize #1}}

\title{Training Large ASR Encoders with Differential Privacy}
\name{Geeticka Chauhan, Steve Chien, Om Thakkar, Abhradeep Thakurta and Arun Narayanan}
\address{Google}
\begin{document}
%
\maketitle
\begin{abstract}
Self-supervised learning (SSL) methods for large speech models have proven to be highly effective at ASR. With the interest in public deployment of large pre-trained models, there is a rising concern for unintended memorization and leakage of sensitive data points from the training data. In this paper, we apply differentially private (DP) pre-training to a SOTA Conformer-based encoder, and study its performance on a downstream ASR task assuming the fine-tuning data is public. This paper is the first to apply DP to SSL for ASR, investigating the DP noise tolerance of the BEST-RQ pre-training method. Notably, we introduce a novel variant of model pruning called \textit{gradient-based layer freezing} that provides strong improvements in privacy-utility-compute trade-offs. Our approach yields a LibriSpeech test-clean/other WER (\%) of 3.78/ 8.41 with ($10$, $\sci{1}{-9}$)-DP for extrapolation towards low dataset scales, and 2.81/ 5.89 with ($10$, $\sci{7.9}{-11}$)-DP for extrapolation towards high scales.
\end{abstract}
%
\section{Introduction}
\label{sec:introduction}

Across various sub-fields of Machine Learning (ML), large scale transformer-based models \cite{vaswani2017attention, zhang2020transformer} have seen widespread adoption for modeling long-range dependencies in sequences. In automatic speech recognition (ASR), the known success of convolutions \cite{li19convacoustic} prompted the introduction of the Conformer architecture \cite{gulati2020conformer}, and later incorporation of BERT-style self-supervised learning (SSL) via the BEST-RQ pre-training method \cite{chiu2022bestrq}. Popular ASR models are often released as modifiable checkpoints after being pre-trained on thousands of hours of crawled user-spoken utterances. Following this paradigm of pre-training ASR encoders on massive amount of data can put the model at risk for leaking sensitive information, especially when the data consist of web crawls that can contain sensitive information such as gender, dialect or identity of a speaker. 

\begin{figure}
     \centering
     \includegraphics[width=\columnwidth]{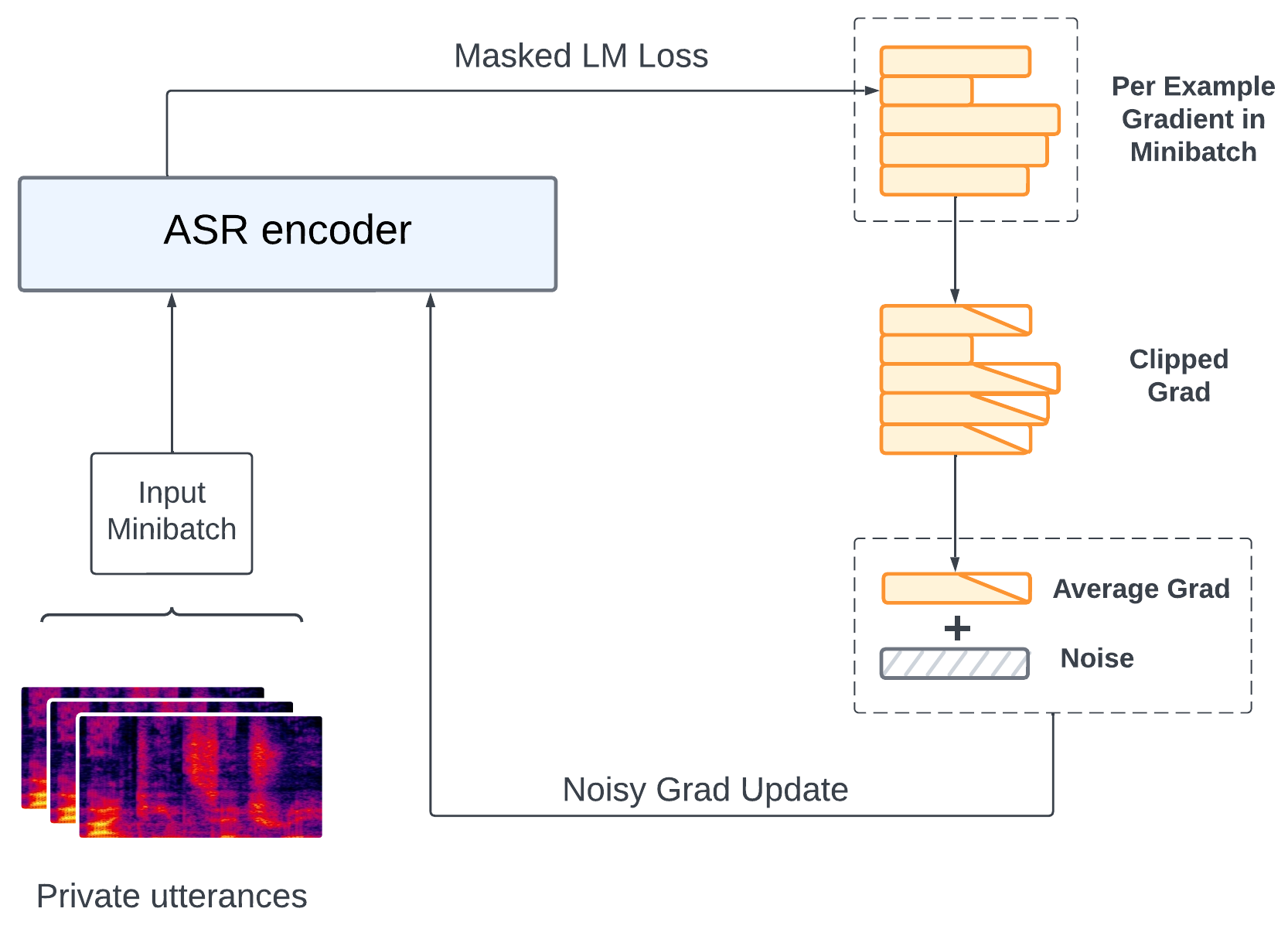}
     \caption{The Differentially Private pre-training method for ASR encoder involving clipping per-example gradients from the minibatch, and addition of calibrated Gaussian noise. Gradients with norms below clip value are not clipped, as shown above. Once private pre-training of the ASR encoder is done, fine-tuning is done publicly after attaching an ASR decoder and using CTC loss \cite{gulati2020conformer, graves2006connectionist}}
     \label{fig:dp-pretrain-figure}
\end{figure}

It is well-known that ML models can leak sensitive information about their training dataset, even when the data is kept private. This has been extensively discussed by works such as \cite{shokri2017membership, carlini2019secret, carlini2021extracting, carlini2023extracting} in natural language processing (NLP) and computer vision (CV)  and later extended to the speech domain by \cite{amid2022extracting, jagielski2022measuring,  wang2023unintended, jagielski2024, shejwalkar24quantifying}.  This paper explores methods focused on differentially private (DP) pre-training of ASR encoders (illustration in Figure \ref{fig:dp-pretrain-figure}) for mitigating the privacy leakage from trained encoders. 

Differential Privacy \cite{dwork2006calibrating} provides a robust way to combat the privacy leakage issue. It provides theoretical guarantees about the limits of influence of any individual training point towards the final model, preventing attackers from confidently inferring whether any particular data sample was used for training. For training large models, DP training is challenging due to the the stringent trade-off between privacy, utility and compute (shortened as trade-offs in this paper). 

A standard mechanism of ensuring DP during model training is the addition of noise to the gradients, which can increase privacy but negatively affect model performance. Increasing batch size \cite{kairouz2021practical, pelikan2023federated} can mitigate this trade-off but increases compute costs. Recent works in language modeling and vision have demonstrated the utility of DP methods being close to their non-private baselines \cite{yu2021differentially, bu2022differentially}. Most works focus on improving trade-offs for fine-tuning, with positive effects seen for parameter efficient techniques such as LoRA \cite{hu2022lora}. LoRA mitigates the issue of growing DP noise magnitude as the model size is increased \cite{bassily2014private, yu2021differentially}. Existing privacy literature lacks comprehensive evaluation of methods for reducing the trade-off between privacy and accuracy in pre-training language models, despite extensive improvement of trade-offs observed for fine-tuning. 


This paper focuses on DP during pre-training (Figure \ref{fig:dp-pretrain-figure}), where new challenges arise from adding substantial DP noise for full model training. Recent work in language modeling \cite{ponomareva2022training} has successfully narrowed the gap between private DP pre-training and the non-private baseline by employing private tokenization and increased compute. More recently, \cite{pelikan2023federated} explored improving trade-offs in the context of DP Federated Learning (FL) for ASR by utilizing per-layer clipping \cite{brendan2018learning}. 


To date, no research has evaluated training with DP in the SSL setting for ASR. This paper makes the following contributions: 1) We are the first to assess the DP noise tolerance for the BEST-RQ setting of a large Conformer model, and 2) We introduce a novel variant of model pruning called gradient-based layer freezing where we determine the model layers to freeze based on a square of gradient analysis. Collectively, our proposed approach achieves significant improvements in utility (Word Error Rate i.e. WER in our case), while maintaining strong privacy guarantees of $\epsilon = 10$.

\section{Background and Related Work}
\subsection{Differential Privacy}
Differential Privacy (DP) \cite{dwork2006calibrating} is widely considered a gold standard for bounding and quantifying the privacy leakage of sensitive data when performing learning tasks. Intuitively, DP prevents an adversary from confidently making any conclusions about whether any particular data was used in training a model, even while having access to the model and arbitrary external side information. The formal definition of DP depends on the notion of neighboring datasets: we will refer to a pair of datasets $D, D' \in \mathcal{D}$ as neighbors if $D'$ can be obtained from $D$ by adding or removing one data sample.

\begin{definition}[($\epsilon$, $\delta)$-DP]
 A (randomized) algorithm $\mathcal{A} : \mathcal{D} \to \Theta$ is ($\epsilon$, $\delta$)-differentially private if for all pairs of neighboring datasets $D, D' \in \mathcal{D}$, and for any $S \subseteq \Theta$ we have,
\begin{equation}
    P[\mathcal{A}(D) \in S] \le exp(\epsilon) \cdot P[\mathcal{A}(D') \in S] + \delta.
\end{equation}
\end{definition}

Typical recommendations for $\epsilon$ and $\delta$ are to be as small as possible, as $\epsilon$ is the multiplicative factor between the probabilities of the two neighboring datasets and $\delta$ is the additive scalar which controls the strength of the relaxation from the stricter $\epsilon$-DP definition \cite{dwork2006calibrating}. The general recommendation in the literature is to choose $\delta \ll \frac{1}{n}$ where $n$ is the number of records in the dataset \cite{dwork2014algorithmic}. \cite{ponomareva2023dp} recommend different tiers for $\epsilon$ values going from strong formal guarantees to reasonable and weak guarantees, where Tier1 $\coloneqq$ $\epsilon \le 1$, Tier2 $\coloneqq$ $\epsilon \le 10$ and Tier 3 $\coloneqq$ $\epsilon > 10$. 

\subsection{Training with Differential Privacy}
DP can be integrated at various stages of the model lifecycle \cite{ponomareva2023dp}, and this paper focuses on applying DP during the model pre-training stage. In this set-up, the pre-training data is kept private, the model is trained using a noise additive technique such as DP-SGD \cite{abadi2016deep} and the model can be released publicly along with its parameter weights for public fine-tuning. Due to the post-processing property of DP, any modifications to the released model (such as public fine-tuning) hold the same theoretical guarantees over the pre-training data.

Typically, differentially private training is performed using variants of DP-SGD \cite{abadi2016deep}, where the main distinctions from non-private training are the clipping of per-example gradients, and the addition of spherical Gaussian noise, as illustrated (for ASR pre-training) by Figure~\ref{fig:dp-pretrain-figure}.  Note that the magnitude of Gaussian noise (called noise multiplier) is directly correlated with the value of $\epsilon$, calculated using the chosen privacy accounting technique such as the one by \cite{abadi2016deep}. This is implemented as a modification to the gradient computation during the optimization step by computing per-example gradients~\cite{subramani2021enabling}, clipping to limit their per-sample sensitivity, and the addition of calibrated Gaussian noise. Therefore, DP training is relatively independent of the exact choice of optimizer. For our experiments, we rely on the Adam optimizer with DP modifications for example-level DP. Training with DP incurs several challenges as a result of clipping and addition of noise, commonly characterized as privacy-utility-compute trade-offs (truncated as trade-offs in this paper). 



\subsection{Related Work}
Many works \cite{bassily2014private, kairouz2021practical, li2022does} have shown that the trade-offs are substantial for training large neural networks with state-of-the-art techniques like DP-SGD \cite{bassily2014private, abadi2016deep}. Consequently, there has been work \cite{abadi2016deep, anil2022large, kurakin2022toward, de2022unlocking} on pre-training using public data for improving the utility of DP-SGD. 
A recent work~\cite{pelikan2023federated} has considered DP training for ASR models, but focusing on the Federated Learning (FL) regime. Additionally, many works \cite{li2021large, yu2021differentially, bu2022differentially} have focused on privately fine-tuning neural networks (focusing largely on vision and language models (LMs)) after pre-training using public data to improve the trade-offs for DP-SGD.  While it is common in literature to treat pre-training data as public, modern large model pre-training can involve sensitive data that is susceptible to be memorized and potentially leaked. There is only one recent work \cite{ponomareva2022training} that studies DP pre-training for LMs, and demonstrates that such models can be fine-tuned to high accuracies on downstream tasks. Related to modifications on bounding sensitivity within a training step, \cite{watson2023inference} have considered the role of gradient clipping and suggest model pruning as a strategy to improve the trade-offs.

\subsection{Challenges}
DP during pre-training remains a relatively unexplored area, and it is unclear whether commonly used fine-tuning techniques directly applied in this context. We devise novel techniques inspired by prior works, and demonstrate their effectiveness during pre-training. The computationally intensive training process, requiring updates to most model parameters, limits quick exploration and prototyping. To address this, we expanded our experimental exploration by evaluating the model continuously during a stable and early pre-trained checkpoint, confirming that comparisons remain valid during later stages of pre-training. This approach enables us to rigorously evaluate early research ideas and maintain a rapid prototyping pace for optimizing the privacy-utility-compute trade-offs during pre-training, contributing to the advancement of privacy-preserving SSL models for ASR.

\section{Experimental Setup}
For our model, we choose the 300M variant \cite{wang2023unintended} of the state-of-the art ASR model architecture, Conformer XL \cite{zhang2022bigssl}. The encoder is pre-trained on LibriLight (LL) \cite{kahn2020librilight} for 1M steps using self-supervised learning via the BERT-based Speech pre-Training
with Random-projection Quantizer (BEST-RQ) \cite{chiu2022bestrq}. Fine-tuning is done for 60k steps post attaching an additional projection layer on the encoder, using the LibriSpeech (LS) \cite{panayotov2015librispeech} dataset. 
Hyperparameter details and model architecture follow the BEST-RQ paper ~\cite{chiu2022bestrq}, and official dataset splits were used for training, validation and hyperparameter tuning. Pre-training takes  \texttildelow 1 week on Dragonfish TPUs with 8x8 topology, fine-tuning takes $\le$ 1 day and original batch size was set at 512. 

\textit{Practically}, DP training involves adding spherical Gaussian noise calculated using popular privacy accounting techniques like \cite{abadi2016deep}. Most related works target Tier 1 or Tier 2 privacy guarantees with $\epsilon \le$ 10 \cite{ponomareva2023dp}. Privacy accounting techniques consider various factors such as target $\epsilon$ and $\delta$, dataset size, minibatch size and training epochs to determine the Gaussian \textbf{noise multiplier} added to the gradients during training. Throughout this paper, we will closely correlate the noise multiplier with our target $\epsilon$ of 10 to demonstrate strong privacy guarantees. 

In this paper, we apply DP to the pre-training stage of our model (with LL), and assume that the fine-tuning dataset (LS) for the downstream ASR task is public. 
Utility is reported as test-clean/other WER on the LS dataset. 
We use the updated moments accountant \cite{abadi2016deep, mtz19} for calculating our privacy guarantees. 
We report experiments with different DP noise multipliers in the range $[\sci{1}{-4}, \sci{1}{-2}]$, since we find that noise multipliers beyond $\sci{1}{-2}$ lead to divergence (more details in Section \ref{sec:noise-tolerance-bestrq}).

Since the trade-offs with large model training can be substantial, we follow the extrapolation strategy similar to recent works~\cite{kairouz2021practical, pelikan2023federated}. We extrapolate the $(\epsilon, \delta)$-DP assuming the training dynamic remains unchanged upon linearly scaling minibatch size and noise multiplier (to maintain the expected signal-to-noise ratio for the gradient update) along with scaling the dataset size (for improved privacy accounting).

\begin{figure}[ht!]
     \centering
     \includegraphics[width=\columnwidth]{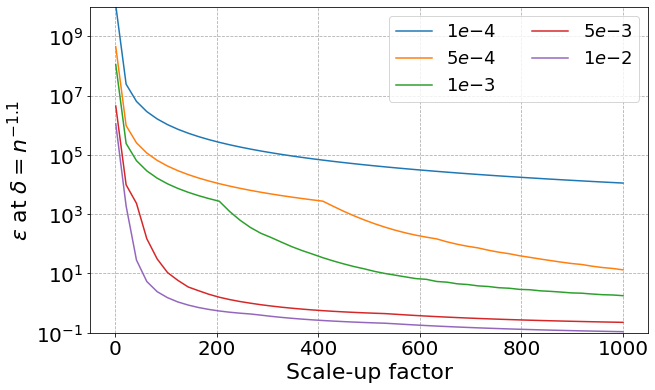}
     \caption{Extrapolating the noise multiplier linearly with batch size and dataset size to maintain the signal-to-noise ratio and improve privacy accounting.}
     \label{fig:dataset-scale-up}
\end{figure}

To evaluate the impact of DP noise multipliers, we experiment with various values and map the corresponding $\epsilon$ using the moments accountant \cite{abadi2016deep, mtz19}. We scale up the batch size, noise multiplier, and pre-training dataset size by a constant factor, as illustrated in Figure \ref{fig:dataset-scale-up}. 
Based on this scaling strategy, we hypothesize that training with a larger, well-curated dataset of the same distribution would yield similar Word Error Rate (WER) performance while improving privacy accounting.
This would allow for a smaller noise multiplier and a stronger $\epsilon$ guarantee.
Figure \ref{fig:dataset-scale-up} illustrates the positive effects of different scale-up factors on $\epsilon$, leading to significant improvements in privacy guarantees. 

Table \ref{tab:scale_up_noise} presents presents the specific scale-up factors for noise multipliers considered in this paper to achieve a DP of $\epsilon = 10$ at $\delta = n^{-1.1}$, where $n$ is the scaled-up dataset size. 
According to recent work~\cite{ponomareva2023dp}, such a level of DP can be classified in the ``Tier 2: Reasonable privacy guarantees". 

\renewcommand{\arraystretch}{1.4}
\begin{table}[th]
    \caption{Extrapolation factor for linearly scaling-up  noise multiplier, batch size and dataset size needed for each used noise multiplier value to get DP $\epsilon = 10$ at $\delta = n^{-1.1}$, where $n$ is the scaled-up dataset size.}
    \centering
    \resizebox{\columnwidth}{!}{
    \begin{tabular}{l l| l l l l l}
    \toprule
         &  \textbf{Noise multiplier}      &$\sci{1}{-4}$   &$\sci{5}{-4}$  &$\sci{1}{-3}$  &$\sci{5}{-3}$  &$\sci{1}{-2}$\\
         \hline
         &  \textbf{Scale-up}    &$5450$ &$1070$ &$530$  &$105$  &$52$   \\
    \bottomrule
    \end{tabular}}
    \label{tab:scale_up_noise}
\end{table}

\section{Results}
In Section~\ref{sec:noise-tolerance-bestrq}, we introduce the baseline of (non-privately) pre-training the BEST-RQ 300M model. We detail preliminary modeling changes required to comply with DP training and include an analysis of the amount of DP noise tolerable for minimal performance regression of the model. Next, in Section~\ref{sec:freeze} following \cite{pelikan2023federated}, we incorporate per layer clipping for improved utility and noise tolerance. Lastly,  we introduce our gradient-based layer freezing strategy (dubbed as LayerFreeze). Our results denote a synergy between per-layer clipping and our model pruning technique, based on the compounding improvements we observe in model quality (summary of results in Table~\ref{tab:improvements}).

\subsection{Noise tolerance of the BEST-RQ 300M model}
\label{sec:noise-tolerance-bestrq}
We establish non-private baselines for the BEST-RQ 300M model, and analyze the degree of noise tolerated for minimal utility regression. As is typical for DP training, we replace batch normalization with group normalization to effectively limit per-sample contributions and avoiding mixing of batch statistics across samples \cite{ponomareva2023dp}. After experimentation, we find the best setting of group normalization to have input rank of 3, number of groups as 1 and group norm epsilon as $\sci{1}{-4}$, resulting in a test-clean/other WER (\%) of 2.17/4.23 post fine-tuning on LibriSpeech. 

Then, we experiment with choices for per-example clipping bounds, and find the bound 1.5 to be clipping almost all samples during training while providing minimal loss in performance, resulting in a WER of 2.21/4.29. We refer to this as the \textbf{non-private lower bound} result. 
Thus, the non-private baseline we report for BEST-RQ consists of group normalization and per-example clipping, to offer a direct comparison to the level of additive noise in our experiments. Our results for the non-private baseline, and for differing level of DP noise are reported in Table \ref{tab:noise-tolerance-initial-model}. 

\begin{table}[th]
  \caption{Noise tolerance of the BEST-RQ 300M model. Our no pre-train upper bound is WER of 4.43/11.23. 
  Above noise multiplier $\sci{1}{-2}$, the model diverges into WER of 100.}
  \label{tab:noise-tolerance-initial-model}
  \centering
  \begin{tabular}{ c c }
    \toprule
    \textbf{Noise multiplier}  &\textbf{test-clean/other WER} \\
    \midrule
    $0$                         &$2.21/4.29$ \\
    $\sci{1}{-4}$                   &$2.24/4.51$ \\
    $\sci{5}{-4}$                   &$2.57/5.98$ \\
    $\sci{1}{-3}$                   &$3.54/8.31$\\
    $\sci{5}{-3}$                   &$10.98/22.62$\\
    $\sci{1}{-2}$                   &$15.38/29.62$\\
    \bottomrule
  \end{tabular}
\end{table}

Note that the performance of the model with fine-tuning from random initialization (no pre-training) is a WER of 4.43/ 11.23, which is the upper bound for effectively measuring the positive effects of pre-training. We refer to this result as the \textbf{no pre-train upper bound}, which is effectively the same as not applying BEST-RQ style pre-training to ConformerXL and just doing supervised training on the Librispeech dataset.
As can be seen from Table~\ref{tab:noise-tolerance-initial-model}, we start seeing significant regressions (greater than 10\% relative) for noise multiplier $\sci{5}{-4}$, where the standard extrapolation technique achieves DP $\epsilon=10$ only at a practically prohibitive scale-up factor of 1070 (Table~\ref{tab:scale_up_noise}).
 For reference, extrapolation factor for DP $\epsilon=10$ from noise multiplier $\sci{1}{-2}$ is as low as 52, though with the current approach we get WER of 15.38/29.62 which is higher than the no pre-train upper bound.
 Our focus in the rest of the paper is to improve trade-offs for the settings with larger noise multipliers in the range $[\sci{5}{-4}, \sci{1}{-2}]$.

\subsection{Improving the noise tolerance via warm-starting, per-layer clipping and gradient-based layer freezing}
\label{sec:freeze}

Recent studies have made significant strides in optimizing DP training, and we incorporated these findings into our experimental design. Ganesh et al. \cite{ganesh2023public} show the importance of public pre-training for private model training, especially with an in-domain public checkpoint. Pelikan et al. \cite{pelikan2023federated} revive per-layer clipping and show improvements for DP in the supervised training setting of FL for ASR. A couple of recent works \cite{luo2021scalable, watson2023inference} have shed light on the benefits of model pruning for DP training, by minimizing the negative effects of compounding noise affected by the model dimensionality.
Thus, in order to bridge the utility gap with the non-private pre-trained baseline, we consider the following three improvements: warm-starting (WS) using public data, per-layer clipping, and our novel method of gradient-based layer freezing. 
Table \ref{tab:improvements} summarizes the compounding improvements on the three considered techniques. 

\renewcommand{\arraystretch}{1.45} 
\begin{table}[th]
    \caption{Final noise tolerance WERs for BEST-RQ 300M model with our considered improvements. If we observe divergence (mainly for higher noise multipliers), we report results on fine-tuning with an early 200k step pre-trained checkpoint instead.
    }
    \label{tab:improvements}
    \centering
    \begin{tabular}{r ccc }
    \toprule
    \textbf{Noise} & \textbf{Public WS} & \textbf{+PerLayerClip} & \textbf{+LayerFreeze} \\
    \hline 
    $\sci{5}{-4}$ & $3.82/7.65$ & $2.78/5.9$ & $\textbf{2.67/5.74}$ \\
    $\sci{1}{-3}$& $4.03/8.62$ & $2.85/6.02$ & $\textbf{2.81/5.89}$ \\
    $\sci{5}{-3}$& $6.34/13.88$ & $3.78/8.09$ & $\textbf{3.19/7.17}$ \\
    $\sci{1}{-2}$& $8.16/17.42$ & $100/100$ & {$\textbf{3.78/8.41}$} \\ 
    \bottomrule
    \end{tabular}
\end{table}

\subsubsection{Warm-starting using in-domain public data (Public WS)}

Following prior work on using in-domain public data for warmstarting DP training \cite{amid2022dpmd, ganesh2023public}, we randomly selected 1\% of the LibriLight (LL) train dataset as a surrogate for a small amount of available in-domain public data.
Further, for improved trade-offs, we conduct the DP pre-training on the entire LL train dataset (i.e., samples in the 1\% public partition are incorporated into the private training dataset, providing a marginal improvement in the privacy accounting).
Fine-tuning with LibriSpeech (LS) after only (non-private) pre-training with 1\% LL  yields a WER of 3.88/8.94. 
Note that this is better than our no pre-train upper bound of 4.43/11.23, but still substantially worse than the non-private lower bound of 2.21/4.29, validating the assumption about only a small amount on in-distribution public data being available in practical scenarios.

We present the results with public warmstart in the second column in Table~\ref{tab:improvements}, and compared to the random initialization results in Table~\ref{tab:noise-tolerance-initial-model}, we observe a slight regression for smaller noise multipliers $\{\sci{5}{-4}, \sci{1}{-3}\}$, whereas a significant improvement for the higher noise multipliers $\{\sci{5}{-3}, \sci{1}{-2}\}$. 

\subsubsection{Per-Layer Clipping}
There are two commonly-used variants of per-layer clipping~\cite{brendan2018learning, pelikan2023federated}, denoted by the \textit{uniform} variant (which splits the clipping bound equally amongst all layers), and the \textit{dim} variant (which splits the clipping bound proportional to each layer's dimension).
We conducted experiments using both the variants, and but found the \textit{dim} variant to be outperforming the uniform one (similar to results seen in \cite{pelikan2023federated}). 

We present the results for adding per-layer clipping for DP pre-training, post public warmstarting, in the third column in Table~\ref{tab:improvements}. While we observe the model diverging for the highest noise multiplier of $\sci{1}{-2}$, we notice significant improvements in model quality for all other considered values of noise multiplier, corroborating the observation in \cite{pelikan2023federated} regarding the usefulness of per-layer clipping in the ASR domain.

\subsubsection{Gradient-based layer freezing (LayerFreeze)}

For reducing the dimensionality of DP training, some recent works~\cite{luo2021scalable, watson2023inference} propose starting from a pruned model that is initialized from a publicly pre-trained checkpoint. 
In this work, we devise a novel one-shot variant of model pruning called \textbf{Gradient-based Layer Freezing} (Algorithm \ref{alg:layer-freeze}), where instead of removing or freezing individual parameters based on their magnitudes, we freeze them layer-wise based on the normalized squared $\boldsymbol\ell_2$ norm of their gradients observed throughout the public warmstarting phase. 
After this operation, we continue DP pre-training with the pruned model and the  entire LL dataset. 

\begin{algorithm}[!ht]
\caption{Gradient-based Layer Freezing (LayerFreeze)}\label{lf-algorithm}
\label{alg:layer-freeze}
\begin{algorithmic}[1]
    \Input{Model $F$ with params $\theta \in \mathbb{R}^M$, num layers $\{i\}_{i=1}^{L}$, Loss fn $\mathcal{L}(\theta)$ over minibatch, Num iterations $T$, Optimizer \texttt{opt}, Grad \texttt{update()} fn, total params per layer \texttt{dim} () fn, top params $p\%$, \texttt{freeze\_top\_layers } whether to freeze top layers or the rest}
        \vspace{0.3em}
        \State $\mathbf{u_o} \gets 0$ \Comment{Init sq grad vector with $M$ dim}
        \For{$t \in [T]$}
            \State $\mathbf{g_t} \gets \nabla_{\theta_t} \mathcal{L}(\theta_t)$
            \State $\mathbf{u_t \gets u_{t-1} + g_t^2}$ \Comment{Accumulate sq grad}
            \State $\theta_t \gets \texttt{update}(\texttt{opt}, \mathbf{g_t}, \theta_{t-1})$
        \EndFor
    
        \For{$i \in [L]$}
            \State $\mathbf{g_{layer_i}} \gets \sum_{\text{param} \in i} \mathbf{u_T}[\texttt{\footnotesize{param}}] / \texttt{dim}(\footnotesize{i})$
        \EndFor

        \State $\footnotesize{top\_par} \gets p \cdot M$ \Comment{Top num of params}
        \State $\footnotesize{top\_layers} \gets [\,]$
        \vspace{0.3em}
        \For{$layer_i, \huge{\_}$ in \texttt{\footnotesize{sorted}} (\texttt{\footnotesize{enumerate}} ($\mathbf{g_{layer}}$))}
        \vspace{0.3em}
        \footnotesize{
            \If{\,$\sum\texttt{dim} (top\_layers +  [layer_i]) \le top\_par$}
            \vspace{0.3em}
                \State $top\_layers.append(layer_i)$
            \vspace{0.3em}
            \Else \, $break$ \Comment{Sort by grad \& get layers till cut off $p$\%}
            \EndIf
            }
        \EndFor

        \If{\texttt{\footnotesize{freeze\_top\_layers}} is \texttt{True}}
            \State $fr\_layers \gets top\_layers$
        \vspace{0.3em}
        \Else \, $fr\_layers \gets \{i\}_{i=1}^{L} - top\_layers$
        \EndIf
    \vspace{0.3em}
    \State \textbf{return} Model $F$ with frozen layers $fr\_layers$
\end{algorithmic}
\end{algorithm}

\begin{figure}[!ht]
     \centering
     \includegraphics[width=\columnwidth]{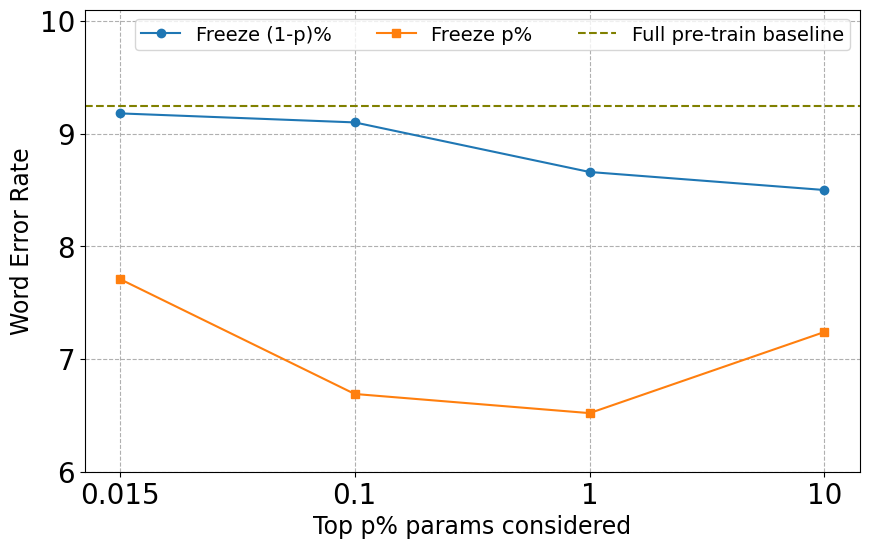}
     \caption{Performance from tuning our LayerFreeze with different percentage of parameters frozen, while keeping the DP noise multiplier constant at $\sci{1}{-3}$. Along the x-axis, we use $p$ to refer to the \% of parameters consisting of layers with the highest accumulated gradient norms. We run experiments with freezing either the $p$\% parameters, or the remaining $(1-p)$\%. To save on compute, fine-tuning is done using an early pre-train checkpoint of 200k, assuming that the same conclusions hold for 1M. }
     \label{fig:dp-layerfreeze}
\end{figure}

Once the norms of the per-layer gradients until our public warmstarting checkpoint are accumulated, we focus on $p$\% of the model parameters, consisting of layers with the highest normalized accumulated squared gradient norm.  We perform tuning experiments by freezing layers associated with either these $p\%$ parameter, or the remaining $1-p\%$ parameters.  
$P$ is treated as a hyperparameter, explored in the range $\{0.015\%,10\%]$ as seen in Figure \ref{fig:dp-layerfreeze}. 
We consistently find that DP pre-training benefits from freezing layers with the top $p\%$ parameters, where the best case is when $p=1\%$. 
To shed additional light on the layers frozen in our best-case scenario, layers corresponding to the bias, scale, beta, and gamma terms were frozen. 
We report the results of using LayerFreeze, along with per-layer clipping and public warmstarting, in the fourth column in Table~\ref{tab:improvements}.
It is important to note that LayerFreeze provides significant improvements in model quality in all the considered settings.

In summary, we obtain LibriSpeech WERs of 3.78/8.41  with ($10$, $\sci{1}{-9}$)-DP for LibriLight with an extrapolation factor of 52 (low dataset scaling regime), and 2.81/5.89 with ($10$, $\sci{7.9}{-11}$)-DP for LibriLight with an extrapolation factor of 530 (high dataset scaling regime).



\section{Conclusion}
We introduce DP to SSL for ASR, and a novel variant of model pruning called gradient-based layer freezing. 
Our technique improves the trade-offs for DP ASR pre-training, over improvements from public warmstarting and per-layer clipping. 
Overall, we demonstrate a DP training method that improves utility significantly while maintaining robust privacy guarantees under various extrapolation factors. 
Though our work provides a way to pre-train ASR encoders with strong DP guarantees, the extrapolations required to reach those guarantees can be limiting in some practical regimes.
Improving computation trade-offs that we incur for reaching strong DP guarantees is an interesting direction we leave for future investigation.

\section{Acknowledgments}
We would like to thank the following collaborators for supporting this work, offering valuable feedback \& helping with quick prototyping of experiments: Lun Wang, Rajiv Mathews, Nanxin Chen, Brennan Saeta, Josh Lipschultz, Qiao Zhang, Colin Gaffney, Virat Shejwalkar and Hongbin Liu.



\bibliographystyle{IEEEbib}
\bibliography{main}
\newpage
\begin{appendices}
\section{Additional Extrapolation Analysis}

\begin{figure}[h]
     \centering
     \begin{subfigure}[b]{0.4\textwidth}
         \centering
         \includegraphics[width=\textwidth]{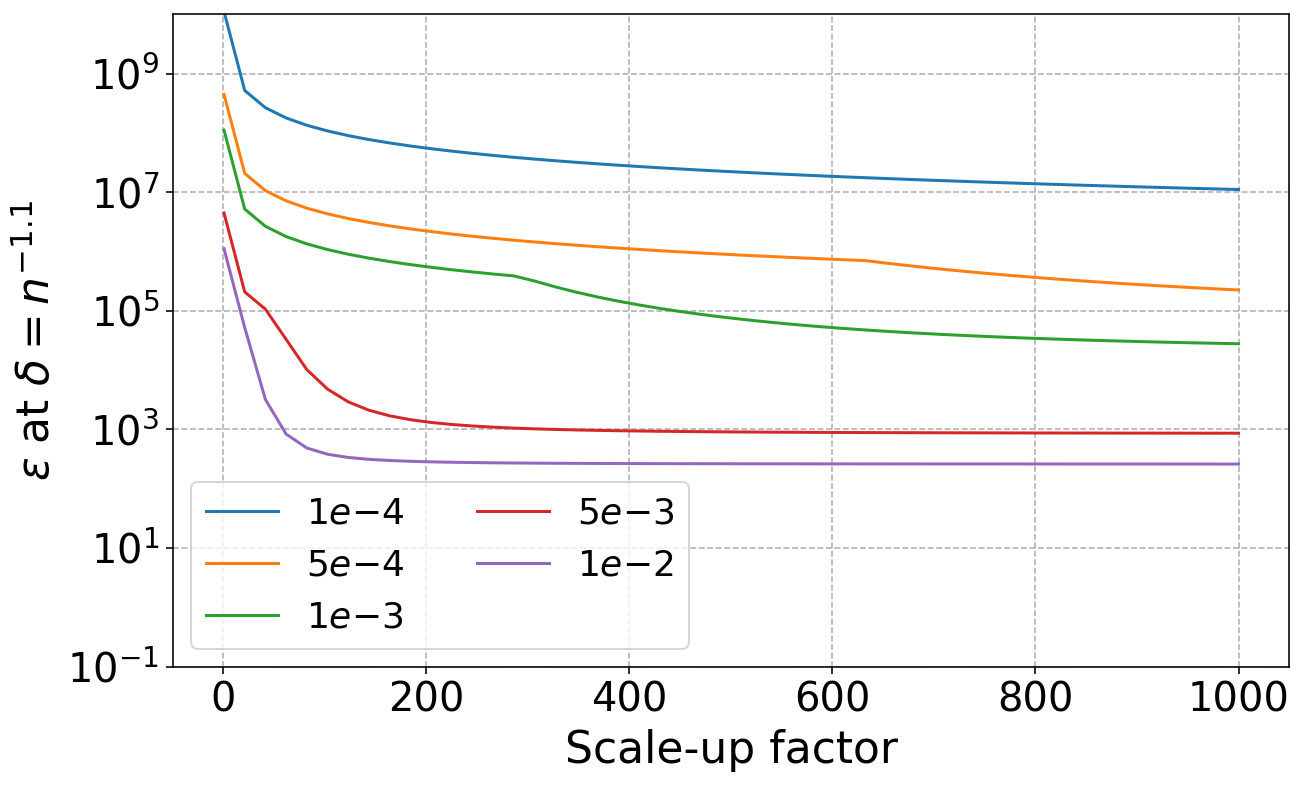}
         \caption{Standard setting: Scaling up the noise multiplier linearly with batch size}
         \label{fig:orig-batch-multiplier}
     \end{subfigure}
     \hfill
     \begin{subfigure}[b]{0.4\textwidth}
         \centering
         \includegraphics[width=\textwidth]{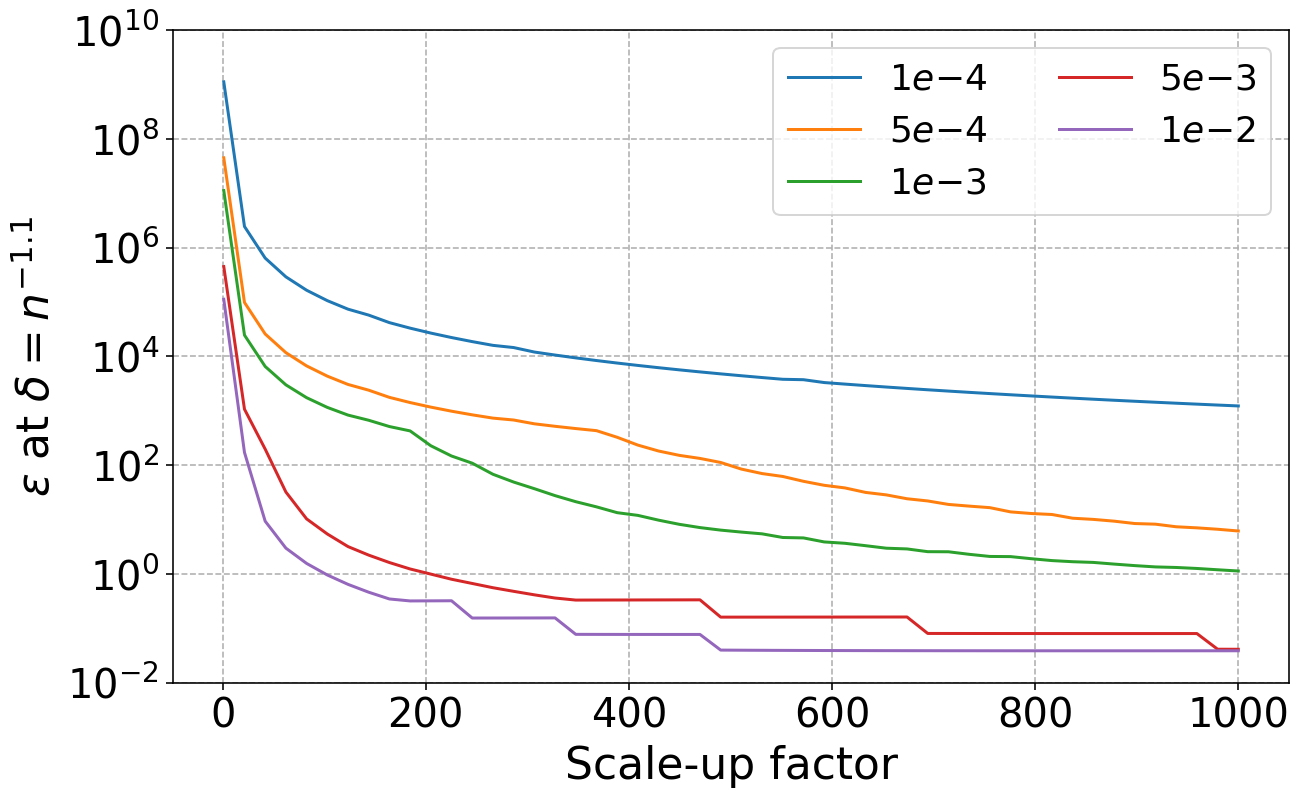}
         \caption{Scaling up the noise multiplier linearly with batch size and dataset size, with the dataset size having a headstart by 10x at each multiplier setting}
         \label{fig:dataset-scale-up10x}
     \end{subfigure}
        \caption{Most extreme setting: Scaling up the noise multiplier linearly with batch size and other independent parameters to maintain the signal to noise ratio. All other training dynamics remain unchanged with the assumption that the utility would remain the same.}
        \label{fig:batch-multiplier-analysis}
\end{figure}

For our extrapolation analysis for noise multiplier, batch size and our batch multiplier analysis, we consider 3 settings: 1) the original setting where only the batch multiplier is scaled up, 2) dataset scale up setting where batch multiplier is scaled up by the same factor as the dataset scale up, and 3) dataset scale up with dataset being 10x higher than the batch scale up. In all settings, we assume batch multipliers beyond 1000 to be too intractable to report. 

We show the effects of settings 1 and 3 in the Figure \ref{fig:batch-multiplier-analysis}, whereas we see the effects of setting 2 in Figure \ref{fig:dataset-scale-up}. We can see that the original setting only allows $\epsilon < 10$ for the noise multiplier 0.1 where the model diverges in utility. However, the more extreme settings in Figures \ref{fig:dataset-scale-up} and \ref{fig:dataset-scale-up10x} allow even noise multiplier of $\sci{1}{-3}$ reach $\epsilon < 10$ for batch multipliers $>$ 400. The ideal setting is with noise $\sci{1}{-2}$ that allows for batch scale up (and corresponding dataset scale ups) for around 50x.



\section{Global and per-layer clip experiments with no noise}
Prior to adding in noise, we experimented with different clipping values, while noting the fraction of clipped gradients. For both global and per-layer clip values, we ensured that the clip values caused minimal loss of utility while clipping the maximum fraction of gradients. It is expected that higher clip values would arrive closer to the no-clip set up, but would not allow us to bind sensitivity for DP training due to the fewer fraction of gradients clipped. Therefore, we only selected both global and per-layer clip values below 5, which would clip most gradients and lead to minimal loss of utility. Based on the results in tables \ref{tab:global-clip} and \ref{tab:per-layer-clip}, we selected clip value 1.5 for the global setting and 0.1 for the per layer clip setting with the \textit{dim} variant. 

\begin{table}[h]
  \caption{Global Clip Result for BEST-RQ 300M model with Group Normalization.}
  \label{tab:global-clip}
  \centering
  \begin{tabular}{ c c }
    \toprule
    \textbf{Clip Value}  &\textbf{test-other WER} \\
    \midrule
    $0.5$                         &$4.55$ \\
    $1$                         &$4.38$ \\
    $\textbf{1.5}$              &$\textbf{4.29}$ \\
    $2.5$                         &$4.53$ \\
    $5$                           &$4.69$\\
    $7.5$                         &$4.88$\\
    $10$                          &$4.47$\\
    $None$                          &$4.23$\\
    \bottomrule
  \end{tabular}
\end{table}

\begin{table}[h]
  \caption{Per Layer Clip Result for BEST-RQ 300M model with Group Normalization. Reporting best intermediate results among the \textit{dim} or \textit{uniform} variant. To save on compute, fine-tuning is done using an early pre-train checkpoint of 200k, assuming that the same conclusions hold for 1M. }
  \label{tab:per-layer-clip}
  \centering
  \begin{tabular}{ c c l }
    \toprule
    \textbf{Clip Value}  &\textbf{test-other WER} &\textbf{uniform/ dim} \\
    \midrule
    $0.01$                         &$9.52$        & \textit{uniform} \\
    $0.03$                         &$7.28$        & \textit{dim} \\
    $0.05$                         &$6.08$        & \textit{dim} \\
    $\textbf{0.1}$             &$\textbf{5.43}$ & \textbf{\textit{dim}} \\
    $0.5$                          &$5.72$       & \textit{uniform} \\
    $1.5$                         &$6.14$       & \textit{uniform} \\
    $2.5$                         &$5.7$        & \textit{uniform} \\
    $5$                         &$5.55$         & \textit{dim} \\
    $7$                         &$5.35$         & \textit{uniform} \\
    $None$                          &$5.32$    & $None$\\
    \bottomrule
  \end{tabular}
\end{table}

\section{Results for tuning LayerFreeze with different percentage of parameters frozen}
In Figure \ref{fig:dp-layerfreeze}, we report intermediate results for freezing different percentage of parameters, ranging from freezing the top $p$\% to the remaining $(1-p)$\%. In table \ref{tab:layer-freeze-ablation-nums}, we report the exact test-other WER corresponding to the figure. 

\begin{table}[th]
  \caption{Tuning our LayerFreeze with different percentage of parameters frozen, while keeping the DP noise multiplier constant at $\sci{1}{-3}$. To save on compute, fine-tuning is done using an early pre-train checkpoint of 200k, assuming that the same conclusions hold for 1M. }
  \label{tab:layer-freeze-ablation-nums}
  \centering
  \begin{tabular}{ c c l }
    \toprule
    \textbf{$P$}  &\textbf{test-other WER} &\textbf{Freeze $P$ or $(1-P)$\%} \\
    \midrule
    -                         &$9.25$        & \textit{No Freezing} \\
    $0.015$                         &$7.71$        & Freeze $P$ \\
    $0.015$                         &$9.18$        & Freeze $1-P$ \\
    $0.1$                     &$6.69$ & Freeze $P$ \\
    $0.1$                          &$9.1$       & Freeze $1-P$ \\
    $1$                         &$6.52$       & Freeze $P$ \\
    $1$                         &$8.66$        & Freeze $1-P$ \\
    $10$                         &$7.24$         & Freeze $P$ \\
    $10$                         &$8.5$         & Freeze $1-P$ \\
    \bottomrule
  \end{tabular}
\end{table}

\end{appendices}

\end{document}